\documentclass[prl,twocolumn,showpacs,preprintnumbers,floatfix,amsmath,amssymb,superscriptaddress]{revtex4}

\usepackage{mathtools}

\usepackage{graphicx}    % Include figure files
\usepackage{dcolumn}     % Align table columns on decimal point<
\usepackage{bm}          % bold math
\usepackage[usenames,dvipsnames]{color}
\usepackage[colorlinks=true ,citecolor=blue]{hyperref}

\newcommand{\ee}{\text{e}}
\newcommand{\ii}{\text{i}}

\begin{document}

\title{Transport properties of  an electron-hole bilayer/superconductor hybrid junction}

\author{D. Bercioux}
\email{dario.bercioux@dipc.org}
\affiliation{Donostia International Physics Center (DIPC), Manuel de Lardizbal 4, E-20018 San Sebasti\'an, Spain}
\affiliation{IKERBASQUE, Basque Foundation of Science, 48011 Bilbao, Basque Country, Spain}
\author{T.M. Klapwijk}
\affiliation{Kavli Institute of Nanoscience Delft University of Technology}
\author{F. S. Bergeret}
\email{sebastian\_bergeret@ehu.eus}
\affiliation{Donostia International Physics Center (DIPC), Manuel de Lardizbal 4, E-20018 San Sebasti\'an, Spain}
\affiliation{Centro de F\'isica de Materiales (CFM-MPC) Centro Mixto CSIC-UPV/EHU,
20018 Donostia-San Sebastian, Basque Country, Spain}

\begin{abstract}
We investigate the transport properties of a junction consisting of an  electron-hole bilayer in contact 
 with  normal and  superconducting leads.  The electron-hole bilayer is  considered as a  semi-metal with  two electronic
  bands.   We assume that in the region between the contacts the system  
  hosts an exciton condensate described by a BCS-like model 
with a gap $\Gamma$ in the quasiparticle density of states.  We  first discuss how 
 the subgap electronic transport through 
the junction is mainly governed by the interplay between two kinds of reflection processes at the interfaces: 
 The standard Andreev reflection at the interface between the superconductor and the exciton condensate, and 
a coherent crossed reflection at the semi-metal/exciton-condensate interface that converts  
electrons from one layer into the other. We show that the differential conductance of the junction shows a 
minimum at voltages of the order of $\Gamma/e$.  Such a minimum can be seen as a direct hallmark of the existence of the gapped
 excitonic state.
\end{abstract}


% 74.45.+c	Proximity effects; Andreev reflection; SN and SNS junctions
% 71.35.-y	Excitons and related phenomena
% 73.20.Mf	Collective excitations (including excitons, polarons, plasmons and other charge-density excitations)

\maketitle

\emph{Introduction.|}  Semimetals (SMs) could undergo at sufficiently low temperatures, a phase transition into an insulating state described by electron-hole bound pairs. These pairs form a so-called exciton condensate (EC), as  theoretically predicted a long time ago, ~\cite{Keldysh:1965, Jerome:1967, Combescot:2016} and one refers to the system being in an excitonic insulating phase. The ground state of such a phase can be described with the help of a BCS-like theory, in analogy to the superconducting phase.  However, the coupling strength in an excitonic insulator is expected to be even weaker than in a superconductor  (S).  Furthermore,  electron-hole recombination can be quite fast, thereby preventing the formation of the condensate. For these reasons the  EC remains an elusive phase of matter~\cite{Keldysh:1965, Jerome:1967, Halperin:1968}. Possible  SM candidates suggested to undergo a transition to the excitonic insulating phase with an EC  are transition-metal dichalcogenide TiSe$_2$~\cite{Monney:2010,Rossnagel:2011}  and  HgTe QWs with a thickness of 20 nm~\cite{Minkov:2013, Knap:2014, Kononov:2016}. However, there  is no conclusive evidence for an EC in such systems.  
 So far the most successful attempt to obtain an EC is based on the condensation of excitons coupled to light confined within CdTe/CdMgTe micro-cavities | the so-called exciton-polaritons~\cite{Kasprzak:2006}.

Besides bulk semi-metals,  there have been several proposals to create an EC in systems with spatially separated electron and hole gases in order to reduce the electron-hole recombination rate.  The exciton formation in such electron-hole bilayers can be detected by Coulomb drag measurements~\cite{Croxall:2008, Seamons:2009, Li:2016, Lee:2016,Fogler:2014,Calman:2016}. According to the theory, if the excitons form a condensate one expects a discontinuity in the drag at the critical temperature and a divergence when $T\rightarrow0$~\cite{Vignale:1996}. Although certain anomalies of the Coulomb drag as a function of temperature have been observed~\cite{Croxall:2008, Seamons:2009},  it is hard to attribute them to the formation of an exciton condensate.

%
%
%%%%%%%%%%%%
\begin{figure}[!t]
\begin{center}
\includegraphics[width=\columnwidth]{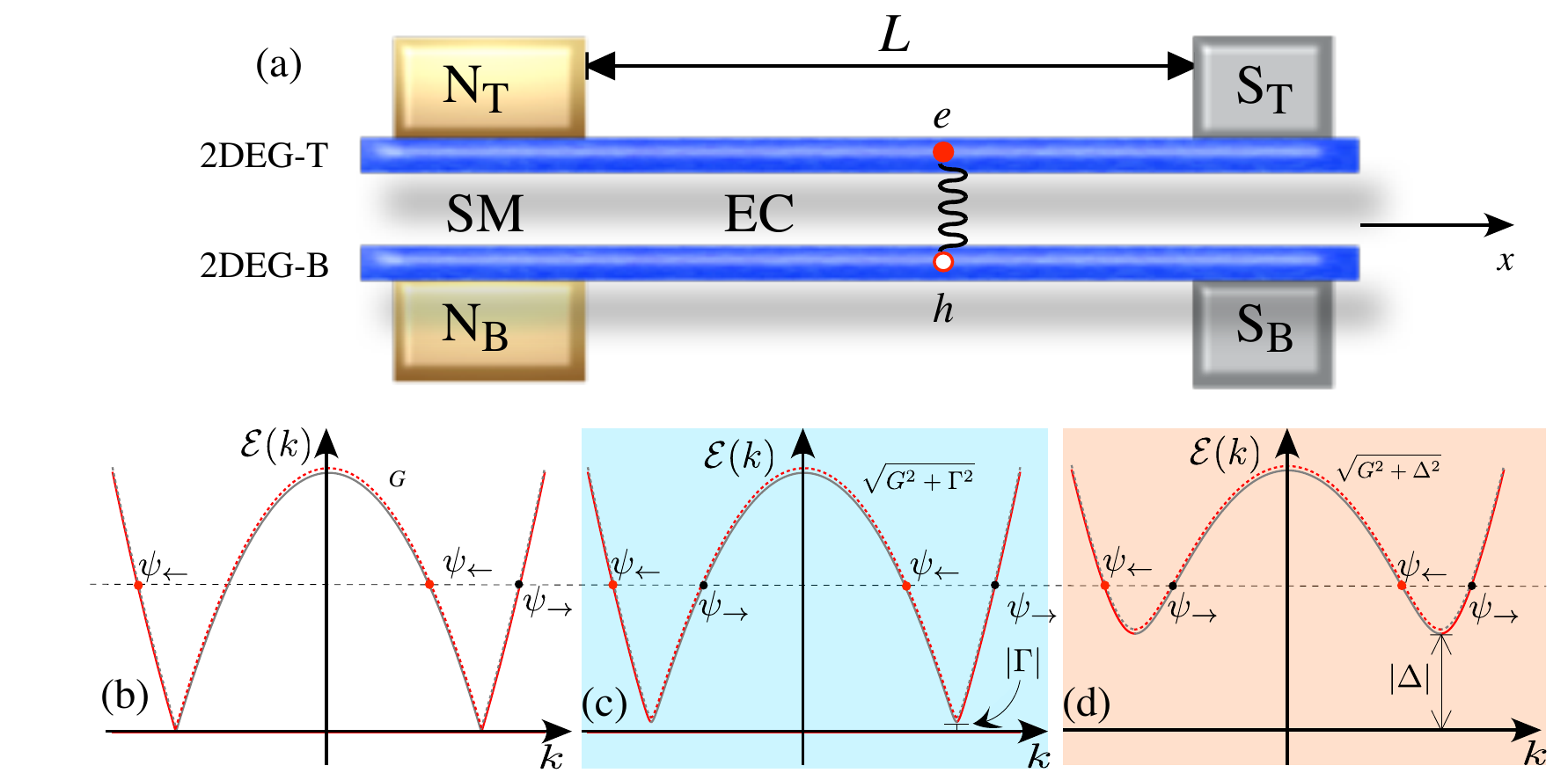}
\caption{\label{figure1} (a) Sketch of the electron-hole bilayer with the two normal (N$_\text{T/B}$) and two superconducting electrodes (S$_\text{T/B}$), the EC region in the middle has length $L$. Energy spectra for the SM region (b), the EC region (c), and SSM one (d).}
\end{center}
\end{figure}
%%%%%%%%%%%%
%
%
%
The primary motivation of this Letter is to propose  and explore  an additional type of measurements 
 to validate the existence of an EC  in electron-hole bilayers. Instead of measuring the Coulomb drag, we suggest to 
 perform a  differential conductance measurement using normal and superconducting electrodes (see Fig.~\ref{figure1}), so to directly unveil the presence of the excitonic gap. 
 We  show that the transport  at low voltages is determined by the competition between the intra-layer Andreev reflection and the inter-layer normal reflection between the  SM and the EC  parts of the system. The latter process is analogous to the one introduced by Rontani and Sham for direct SM/EC interfaces~\cite{Rontani:2005, Rontani:2005b,Wang:2005}.   We discuss the competition between these two type of  reflections as a function of the length of the EC region and show that this competition leads to a minimum of the full differential conductance at a  voltage $V_0$ of the order of the EC order parameter.

\emph{Model and Formalism.|} We consider an ideal two-dimensional electron-hole bilayer, characterized by two parallel two-dimensional electron gases with opposite particle filling, as illustrated in Fig.~\ref{figure1}(a) | the top layer (TL) is ``electron-doped" whereas the bottom layer (BL) is ``hole-doped".  By gating the two layers independently  it is possible to  modulate the band overlap $G$ between the bands of the TL/BL~\cite{Zhu:1995}.
We assume a spatial  modulation along the $x$-axes of the Coulomb interaction in the electron-hole bilayer. In the parts of the bilayer covered by the metallic electrodes, the charge screening allows neglecting the Coulomb interaction. Thus the left and right parts of the system,  are described by a SM consisting of two bands | one from each layer | crossing at the Fermi level [see Fig.~\ref{figure1}(a)].   Due to the proximity effect, we assume an induced superconducting gap $\Delta$ on the part of the layers below the S contacts [see Fig.~\ref{figure1}(c)]. In the region between the contacts the reduced screening results in an \emph{indirect} EC described by the order parameter $\Gamma$~\cite{Zhu:1995} [see Fig.~\ref{figure1}(b)].

To be precise, the structure shown in Fig.~\ref{figure1}(a) is modelled as a  SM, \emph{i.e.}, $\Gamma=\Delta=0$, in contact with  a central region of length $L$ with a finite EC coupling ($\Gamma\neq0$ and $\Delta=0$). In the  region $x>L$   we assume $\Delta\neq0$  due to  the  proximity  from the  S electrodes~\cite{Beenakker:2006} and $\Gamma=0$. We denote this region as SSM.
 This junction is described by the following Hamiltonian written  in an extended Nambu space~\cite{Dolcini:2010}:
%
%
%%%%%%%%%%%%
\begin{align}\label{Ham}
\mathcal{H}_\text{Hyb}= &
\begin{pmatrix}
\frac{\bm{p}^2}{2m}-G & \Gamma(x) & \Delta(x) \ee^{\ii \phi} & 0 \\
\Gamma(x)^* & G - \frac{\bm{p}^2}{2m}  & 0 & \Delta(x) \ee^{\ii \phi} \\
\Delta(x) \ee^{-\ii \phi} & 0 & G-\frac{\bm{p}^2}{2m} & -\Gamma(x) \\
0 & \Delta(x) \ee^{-\ii \phi} & -\Gamma(x)^* &  \frac{\bm{p}^2}{2m}-G
\end{pmatrix}\,,
\end{align}
%%%%%%%%%%%%
%
%
where
\begin{subequations}
\begin{align}
\Gamma(x)&=\Gamma\Theta(x)\big(1-\Theta(x-L)\big)\,,\label{Gamma}\\
\Delta(x)&= \Delta \Theta(x-L)\label{Delta}\,,
\end{align}
\end{subequations}
and  $\Theta(x)$ is the Heaviside step function.
The Hamiltonian (\ref{Ham}) is written in the basis defined by the \emph{bi}-spinor  $\Psi=~(\psi_{\text{TL},\sigma},\psi_{\text{BL},\sigma},\psi^\dag_{\text{TL},-\sigma},\psi^\dag_{\text{BL},-\sigma})$, where   $\sigma$ is the spin~\cite{Dolcini:2010,Peotta:2011}. 
We account for possible elastic reflection at the SM/EC and EC/SSM interfaces by introducing delta-barriers in the system Hamiltonian~\cite{Blonder:1982}, 
$\mathcal{H}_\text{int}=H_\text{SM/EC}\delta(x)+H_\text{EC/SSM}\delta(x-L)$.
This reflection can be ascribed, for example,  to the mismatch of the Fermi wave vector in the  different regions~\cite{Imputirty}.

We analyse the scattering properties of this hybrid SM/EC/SSM junction by matching the scattering states at each interface separating these three regions~\cite{SuMa}.
In the SM region, there is no coupling between the TL and BL nor between electron/hole of the same layer. 
Thus, we can consider an electron (hole) in the TL (BL) as an initial scattering state. 
In the middle, EC,  region, there is a finite coupling between electrons of TL and BL proportional to $\Gamma$.
And in the SSM region  the superconducting pairing $\Delta$ couples electrons and holes within the same layer. 

We summarize   all possible processes that  an  incoming electron from the TL of the SM  at energies smaller than $\Gamma$ and $\Delta$,   may experience. 
If we look at the SM/EC interface the electron  can be  either  normal reflected  in the same layer with amplitude | $r_\text{N,T$\to$T}\equiv r_\text{NTT}$, or in the opposite layer | $r_\text{N,T$\to$B}\equiv r_\text{NTB}$. The latter process resembles the Andreev reflection at the S/N interface and was studied by Rontani and Sham in  Ref.~\cite{Rontani:2005b,Rontani:2005,Wang:2005}. It  results in an exciton in the EC region. 
Whereas at the EC/SSM interface the electron  can be   Andreev-reflected   into a  hole either in the same  layer | $r_\text{A,T$\to$T}\equiv r_\text{ATT}$, 
or   into the  other layer | $r_\text{A,T$\to$B}\equiv r_\text{ATB}$.  It is interesting to notice  that  the reflection amplitude $r_\text{ATB}$ is the analog to the  Andreev specular-reflection in Dirac-material/superconductor hybrid junctions~\cite{Beenakker:2006}.

For energies higher than  the EC gap, $\mathcal{E}>|\Gamma|$, the particles travel  through  the EC region as propagating waves. 
In contrast, for $\mathcal{E}<|\Gamma|$,  the quasi-bound states  are characterized by complex momenta  describing evanescent modes. 
The characteristic decay length of these  modes $\xi_\Gamma$, is energy dependent  and given by:
%
%
%%%%%%%%%%%%
\begin{align}\label{xiGamma}
\xi_\Gamma^{-1}=\sqrt{\frac{1}{2}\left(\chi_Q+\sqrt{\chi_Q^2+\frac{4m^2}{\hbar^4}(\Gamma^2-\mathcal{E}^2)}\right)}.
\end{align}
%%%%%%%%%%%%
%
% 
where $\chi_Q=Q^2-\frac{2mG}{\hbar^2}$, here $Q$ is the momentum parallel to the interfaces that is conserved in the scattering process~\cite{SuMa}.  
When the band overlap $G$ is the dominant energy scale, \emph{i.e.}, $G\gg\max[\Gamma,\Delta,\mathcal{E}]$, the EC characteristic length scale is  approximated by 
%
%
%%%%%%%%%%%%
\begin{align}\label{xiGammaApprox}
\xi_\Gamma=\frac{4\hbar\sqrt{mG^3}}{\Gamma(4mG+Q^2\hbar^2)}. 
\end{align}
%%%%%%%%%%%%
%
%
In what  follows, we focus on the subgap transport,  \emph{i.e.} we consider the injection of  electrons from the TL  with energies smaller than  the SSM gap, $\mathcal{E}<|\Delta|$. Furthermore, we also assume that $\Gamma<\Delta<G$.
The probability for the four possible reflection channels in the SM electrode are then given by~\cite{SuMa}:
%
%
%%%%%%%%%%%%
\begin{subequations}\label{probabilities}
\begin{align}
R_\text{NTT}(\alpha)& =|r_\text{NTT}(\alpha)|^2\,, \\
R_\text{NTB}(\alpha)& = |r_\text{NTB}(\alpha)|^2 \left| \frac{\text{Im}[\kappa_-]}{\kappa_+}(\alpha)\right|\Theta(G-\mathcal{E}) \label{NRTB}\,,\\
R_\text{ATT}(\alpha)& = |r_\text{ATT}(\alpha)|^2 \left| \frac{\text{Im}[\kappa_-]}{\kappa_+}(\alpha)\right|\Theta(G-\mathcal{E}) \label{ARTT}\,,\\
R_\text{ATB}(\alpha)& =|r_\text{ATB}(\alpha)|^2\,,
\end{align}
\end{subequations}
%%%%%%%%%%%%
%
%
where $\alpha$ is the injection angle. For  angles larger than a critical  $\alpha_\text{c}=\arcsin[\pm\sqrt{{\mid G-\mathcal{E}\mid}/{\mathcal{E}+G} }]$, $R_\text{NTB}=R_\text{ATT}=0$ because of the lack of propagating states on the TL/BL.
The actual form of the reflection and transmissions amplitudes  are obtained by solving the scattering problem at the two interfaces~\cite{SuMa}.

\emph{Results.|} We now analyze   the dependence of the  probabilities \eqref{probabilities} on  the length of the EC region.
In  the limiting  case  $L=0$ the system consists  of a simple  SM/SSM junction. 
 In this case, the absence of the EC leads to $r_\text{NTB}=0$. 
Moreover, $r_\text{ATB}$ also vanishes and the results for a clean interface coincide with the standard N/S junction case~\cite{Blonder:1982,SuMa}. Here we  stress the analogy between the interlayer Andreev reflection and the specular Andreev reflection typical of SM with Dirac spectrum again~\cite{Beenakker:2006,Efetov:2015,Chen:2013}.  If there is no EC  then electrons from the TL and BL are decoupled, and hence all reflections  occur  within the same layer only.

%
%
%%%%%%%%%%%%
\begin{figure}[!t]
\begin{center}
\includegraphics[width=\columnwidth]{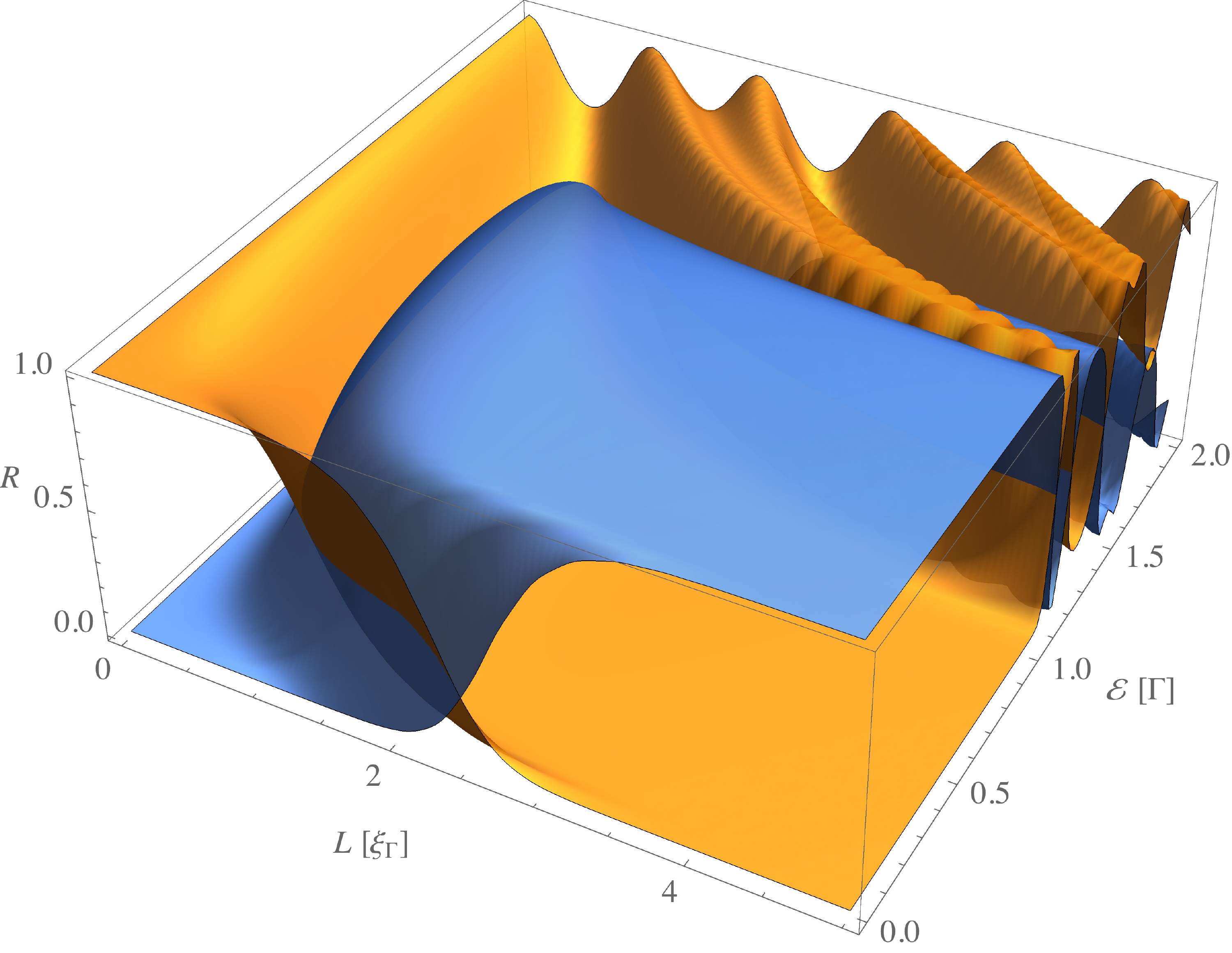}
\caption{\label{figure2} $R_\text{NTB}$ reflection (blue surface) and $R_\text{ATT}$ (orange surface) reflection as a function of the length of  EC region $L$ and  injection energy $\mathcal{E}$, both surfaces are shown for injection energies up to $\Delta$. Here, we have used $\Gamma=1$, $\Delta=2\Gamma$, $G=100\Gamma$, and normal incidence $\alpha=0$.}
\end{center}
\end{figure}
%%%%%%%%%%%%
%
%

We now  focus on the more interesting case of a finite EC region, $L\neq0$.
 In Fig.~\ref{figure2} we present the reflections $R_\text{NTB}$ and $R_\text{ATT}$  as a function of the injection energy 
 and length of the EC region for a normal injection angle ($\alpha=0$).
For a length smaller than $\xi_\Gamma$, the probability for an injected electron in  the TL  to reach the EC/SSM 
interface is large. The  electron is  then Andreev reflected  as a hole of the same layer.
This  explains that for small $L$,  the  Andreev reflection, $R_\text{ATT}$, dominates over the interlayer one,  $R_\text{NTB}$, 
 at all energies.

By increasing the length of the EC region, the probability to reach the EC/SSM interface, for an injected electron with energy $\mathcal{E}<\Gamma$ decreases and hence the $R_\text{ATT}$ processes are suppressed, whereas the  $R_\text{NTB}$  ones are enhanced. The crossover between these two behaviors depends on the injection energy.  In the limiting case  $L\gg\xi_\Gamma$ the probability for the injected  TL electron of reaching the SSM electrode is exponentially small and the  system behaves as a SM/EC junction with the reflection probability $R_\text{NTB}=1$~\cite{Rontani:2005,Rontani:2005b}. For injection energies larger than the EC gap $|\Gamma|$ (but smaller than $\Delta$),  all electrons propagates towards the EC/SSM interface and hence Andreev processes dominate, whereas $R_\text{NTB}$ decreases to zero by increasing the injection energy.  The oscillatory behavior  we observe for this energy range is a feature of quasi-bound states  in the EC region.

Analytically we can calculate the reflection probabilities~\eqref{NRTB} and \eqref{ARTT} in the short-junction limit, $\delta=L/\xi_\Gamma\ll1$  within the Andreev approximation,  {\it i.e.} $G\gg\text{Max}[\mathcal{E},\Delta,\Gamma]$.  In this limits and for zero injection angle $\alpha=0$, we obtain
%
%
%%%%%%%%%%%%
\begin{subequations}\label{seba:an}
\begin{align}
R_\text{NTB}&=4\delta^{2}\frac{\mathcal{E}^{2}\Gamma^{2}}{\Delta^{2}(\Gamma^{2}-\mathcal{E}^{2})}\,,\label{an:RRS} \\
R_\text{ATT}&=1-4\delta^{2}\frac{\mathcal{E}^{2}\Gamma^{2}}{\Delta^{2}(\Gamma^{2}-\mathcal{E}^{2})}\,,\label{an:RAR} \\
R_\text{NTT}&= 0\,.
\end{align}
\end{subequations}
%%%%%%%%%%%%
%
%

\noindent These expressions are in agreement  with the numerical results shown in  Fig.~\ref{figure2}  within the small $L$ region. Notice that for the large value of $G$ chosen, $G=100\Gamma$,   the probability for the  reflection $R_\text{ATB}$ is negligibly small. However, if one choose a smaller $G$ and a larger injection angle $\alpha$, $R_\text{ATB}$ is finite~\cite{SuMa}.  In this latter case  both types of Andreev reflections, $R_\text{ATB}$ and $R_\text{ATT}$, may take place simultaneously. This differs  from the case of retro- and specular-Andreev  reflections  in SM with Dirac spectrum, where one or the other is finite  by crossing the charge neutrality point~\cite{Beenakker:2006,Efetov:2015,Chen:2013}.

To make a  connection with possible transport experiments, we now turn the  focus upon the differential conductance (DC), which is 
a quantity accessible in conventional transport experiments. 
For this purpose, we assume that the two left normal contacts are placed at the same potential $V$ and that the two right superconducting contacts are grounded.  The DC can be expressed in terms of the reflection probabilities~\eqref{probabilities}, by using a generalized expression based on the Blonder-Tinkham-Klapwijk  formula~\cite{Blonder:1982,Mortensen:1999}, 
%
% 
%%%%%%%%%%%%
\begin{align}\label{diffCond}
\frac{\partial I}{\partial V}  &=G_0(eV) \int_{0}^{\frac{\pi}{2}} \sum_{\beta\in\{\text{T,B}\}}\Big[1-R_{\text{NT}\beta}(eV,\alpha)\\
& \hspace{1.5cm} +R_{\text{AT}\beta}(eV,\alpha)\Big]\cos\alpha d\alpha\,.\nonumber
\end{align}
%%%%%%%%%%%%
%
%
Here $G_0(eV)$ is the differential conductance of the normal state.  
%
%
%%%%%%%%%%%%
\begin{figure}[!t]
\begin{center}
\includegraphics[width=\columnwidth]{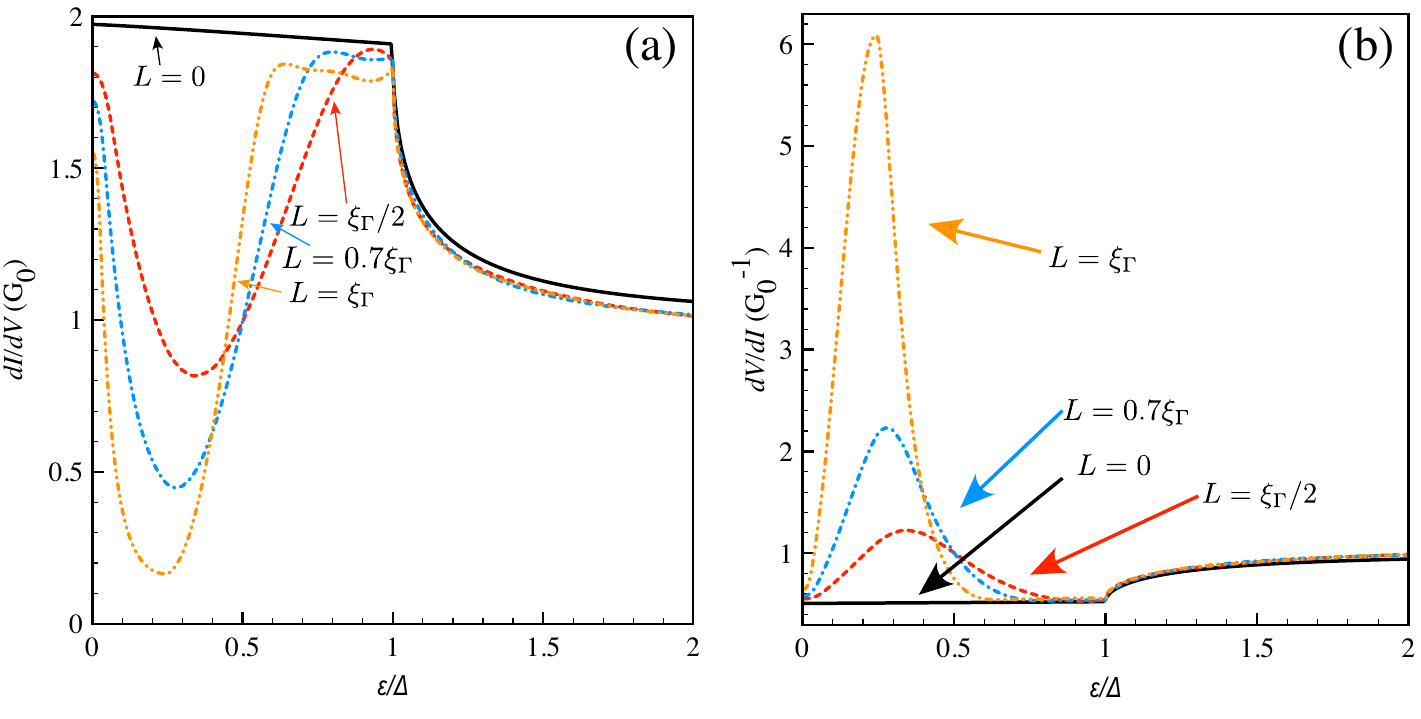}
\caption{\label{figure3} Panel (a): Differential conductance as a function of the applied bias and for different lengths $L$ of the EC region, the various lines correspond to $\Gamma=1$, $\Delta=4\Gamma$,  $G=100\Gamma$ and $H_\text{L}=H_\text{R}=0$. Panel (b): Differential resistance as a function of the applied bias and different length of the EC region. In the figure we have set  $\Gamma=1$, $\Delta=12\Gamma$,  $G=60\Gamma$ and $H_\text{SM/EC}=H_\text{EC/SSM}=0.5$, The parameters are chosen in accordance  to Ref.~[\onlinecite{Kvon:2011,Kononov:2016}]. In both panels The different lines refer to $L=0$ (black-solid line), $L=\xi_\Gamma/2$ (red dashed line), $L=0.7\xi_\Gamma$ (blue dashed-dotted line), and $L=\xi_\Gamma$ (orange dashed-dashed-dotted-dotted line).}
\end{center}
\end{figure}
%%%%%%%%%%%%
%
%

Figure~\ref{figure3}(a) shows the DC  as a function of the injection energy,  which is proportional to the applied voltage, and for different values of the length $L$. Notice first that the DC is not equal to $2G_0$ for $L=0$. This is a consequence of considering a finite chemical potential~\cite{Beenakker:2006}. For a finite length EC region the DC decreases faster as a function of the voltage, it  reaches a minimum and then increases up to a voltage of the order of $\Delta$.  It is also worth to notice that the DC of the normal state in the presence of a finite EC is also smaller compared to the case without it.  

The decrease of the DC is due to a normal reflection channel that accounts for electrons injected from the TL and reflected  into the BL. Thus, the presence of a finite length EC region entirely accounts for the minimum in the DC.
If the  length $L$ exceeds the decay length $\xi_\Gamma$,   injected electrons with $E<\Gamma$ rarely reach  the superconducting electrode (c.f.~Fig.~\ref{figure2}) and therefore  the DC remains  small up to voltages of  the order $|\Gamma|$~\cite{Rontani:2005}.  \\

We can obtain an analytical expression for the low  bias behavior of the DC in the short-junction limit, using Eqs.~\eqref{seba:an}, and within the Andreev approximation:
%
%
%%%%%%%%%%%%
\begin{equation}\label{an:DC}
\frac{dI}{dV} = G_0\left[ 1- 8 \delta^2 \frac{(eV)^2\Gamma^2}{\Delta^2(\Gamma^2-(eV)^2)} \right]\,.
\end{equation}
%%%%%%%%%%%%
%
%
According to this expression the low voltage  peak in the DC is of the order of $\Gamma$. This  suggests that transport measurement using superconducting electrodes could be used to directly estimate the size of the order parameter $\Gamma$.

Our results for the differential conductance may help to understand measurements of the differential resistance of a superconductor/HgTe-quantum well junction. The quantum well has a width of 20 nm~\cite{Kononov:2016}. One of the most striking findings in this experiment is a zero-bias peak in the measured differential resistance~\cite{Kononov:2016} that can be seen as the manifestation of an EC gap. Indeed, in the light of our model the size of the zero-bias peak corresponds to the EC gap. This comparison is shown in Fig. \ref{figure3}(b) where we plot the differential resistance by choosing parameters consistent with the experiments on HgTe quantum wells~\cite{Kvon:2011,Kononov:2016}. Although the agreement between theory and experiment looks promising we would like to be cautious at this point because our approach assumes a full ballistic system, whereas the samples measured in Ref.~\cite{Kononov:2016} have a size of several $\mu$m, and disorder might play a role~\cite{SuMa2}. Further experiments and analysis are needed to draw definite conclusions.

%Our results for the differential conductance may help to understand measurements of the differential resistance of a superconductor/HgTe-quantum well junction. The quantum well has a  width of 20 nm~\cite{Kononov:2016}.  One of the most striking findings in this experiment is  a zero-bias peak in the measured differential resistance~\cite{Kononov:2016} that can be seen as the manifestation of an EC gap. Indeed, in the light of our model the size of the zero-bias peak corresponds to the EC gap.  This comparison is shown in Fig.~\ref{figure3}(b) where we plot the differential resistance by choosing parameters consistently with the experiments on HgTe quantum wells~\cite{Kvon:2011,Kononov:2016}.  Although the agreement between theory and experiment looks promising one still should take cautious at this point because our approach assumes a full ballistic system, whereas the samples measured in Ref.~\cite{Kononov:2016} has a size of several $\mu$m, and disorder might play a role~\cite{SuMa2}. Further experiments are needed to draw final conclusions.

%

\emph{Conclusions.|}  We have studied the electronic transport through an electron-hole bilayer in contact with normal and superconducting electrodes. We have assumed that the electron-hole bilayer  hosts an exciton condensate. 
The transport properties of this junctions are determined by the competition of different coherent reflection processes occurring at the interfaces with the normal and superconducting electrodes. As a consequence of this competition, the differential conductance has a minimum at voltages of the order $\Gamma/e$, where $\Gamma$ is the EC order parameter. 
The observation of this minimum in an electron-hole bilayer system represents a unique hallmark of the presence of the EC.
Good candidates for our proposal are double bilayer-graphene systems separated  by hexagonal boron nitrate~\cite{Li:2016, Lee:2016}.  In bilayer-graphene, superconductivity induced by proximity effect has been already observed~\cite{Efetov:2015}. Moreover,  from our model one might interpret the zero bias-peak observed in the differential resistance of a  HgTe quantum well/superconductor junction as a manifestation of an EC phase.

%\begin{acknowledgments}
\emph{Acknowledgments.}| Discussions with B. Bujnowski, E. Deviatov, F. Konschelle, V. Golovach,  T. T. Heikkil\"a,  P. Lucignano, M. Rontani and T. van den Berg are acknowledged. 
The work of DB and FSB is supported by Spanish Ministerio de Econom\'ia y Competitividad (MINECO) through the project  FIS2014-55987-P and by the Transnational Common Laboratory \emph{QuantumChemPhys}. TMK acknowledges the financial support from the European Research Council Advanced grant no. 339306 (METIQUM), and the Ministry of Education and Science of the Russian Federation, contract N 14.B25.31.0007 of 26 June 2013. 
%\end{acknowledgments}

\bibliography{biblio}

\begin{thebibliography}{33}
\bibitem[{Imp()}]{Imputirty}
\bibinfo{note}{The use of the BTK formalism assumes a sharp boundary between
  the different regions. In particular, an abrupt change of the superconducting
  pair correlations at the SSM/EC interface. This assumption is well justified
  if the size of the contact between the two regions is much smaller than its
  lateral dimensions. This condition can be achieved experimentally by proper
  design of the contact. However, even in the case of a smooth transition of
  the pair correlations at the interface, we expect to have qualitatively
  similar results.}
\end{thebibliography}

\clearpage

\appendix
\section{The scattering states}
In this section we briefly introduce the scattering states in the three regions of the electron-hole bilayer system we have considered in the main text, this is constituted by a SM region on the left, an EC region of length L in the center and a superconducting region on the right.
The scattering states in the SM region read: 
%
%
%%%%%%%%%%%%
\begin{align}\label{SM:state}
\psi_\text{SM}&=\ee^{\ii Q y}\!\!\left\{\left[\ee^{\ii  \kappa_+ x}+ r_\text{NTT}\ee^{-\ii  \kappa_+ x}\right]
\!\!\left(\begin{smallmatrix} 1 \\ 0 \\0 \\0 \end{smallmatrix}\right) \!\! + r_\text{NTB} \ee^{\ii  \kappa_- x}\left(\begin{smallmatrix} 0 \\ 1\\ 0 \\ 0 \end{smallmatrix}\right) \right. \nonumber \\
&+ \left.r_\text{ATT}\ee^{\ii \kappa_- x} \left(\begin{smallmatrix} 0 \\ 0\\ 1 \\ 0 \end{smallmatrix}\right)+ r_\text{ATB} \ee^{-\ii \kappa_+ x} \left(\begin{smallmatrix} 0 \\ 0 \\ 0 \\ 1 \end{smallmatrix}\right)\right\}\,,
\end{align}
%%%%%%%%%%%%
%
%
here $Q$ is the component of the momentum along the $y$ direction. Being that we consider a translational invariant interface between the three regions, the component of the momentum parallel to this interface is a conserved quantity in the overall  scattering process.  At a fixed injection energy $\mathcal{E}$, the momenta $\kappa_\pm$ are expressed by the following relations [c.f. Fig.~1(b) main text]:
%
%
%%%%%%%%%%%%
\begin{equation}\label{momenta:SM}
\kappa_\pm=\sqrt{\frac{2m}{\hbar^2}(G\pm\mathcal{E})-Q^2}\,.
\end{equation}
%%%%%%%%%%%%
%
%

\hspace{2cm}

In the SSM region, the superconducting pairing couples electrons and holes of the same layer, but there is no direct coupling between electrons of the two layers. The incoming electron from the SM region can be transmitted as quasi-electron or quasi-hole either in the top later (TL) ($t_\text{QET}$ and $t_\text{QHT}$)  or bottom layer (BL)  ($t_\text{QEB}$ and $t_\text{QHB}$). The transmitted wave function reads:
%
%
%%%%%%%%%%%%
\begin{align}\label{states:SSM}
\psi_\text{SSM}& = \ee^{\ii Qy} \left[t_\text{QET} \ee^{\ii k_+ x} 
\left(\begin{smallmatrix}
u_\text{T} \\ 0 \\ v_\text{T} \ee^{\ii \phi}\\ 0
\end{smallmatrix}\right)
+ t_\text{QHT} \ee^{-\ii k_- x} 
\left(\begin{smallmatrix} 
v_\text{T} \\ 0 \\ u_\text{T}  \ee^{\ii \phi}\\ 0 
\end{smallmatrix}\right)\right.\nonumber \\
&\left.+t_\text{QHB} \ee^{\ii k_+x}
\left(\begin{smallmatrix}
0 \\ u_\text{B} \\ 0 \\ v_\text{B} \ee^{\ii \phi}
\end{smallmatrix}\right)
+t_\text{QEB} \ee^{-\ii k_- x}
\left(\begin{smallmatrix}
0 \\ v_\text{B} \\ 0 \\ u_\text{B} \ee^{\ii \phi}
\end{smallmatrix}\right)
\right]\,.
\end{align}
%%%%%%%%%%%%
%
%

\begin{figure*}[!t]
\begin{center}
\includegraphics[width=0.9\textwidth]{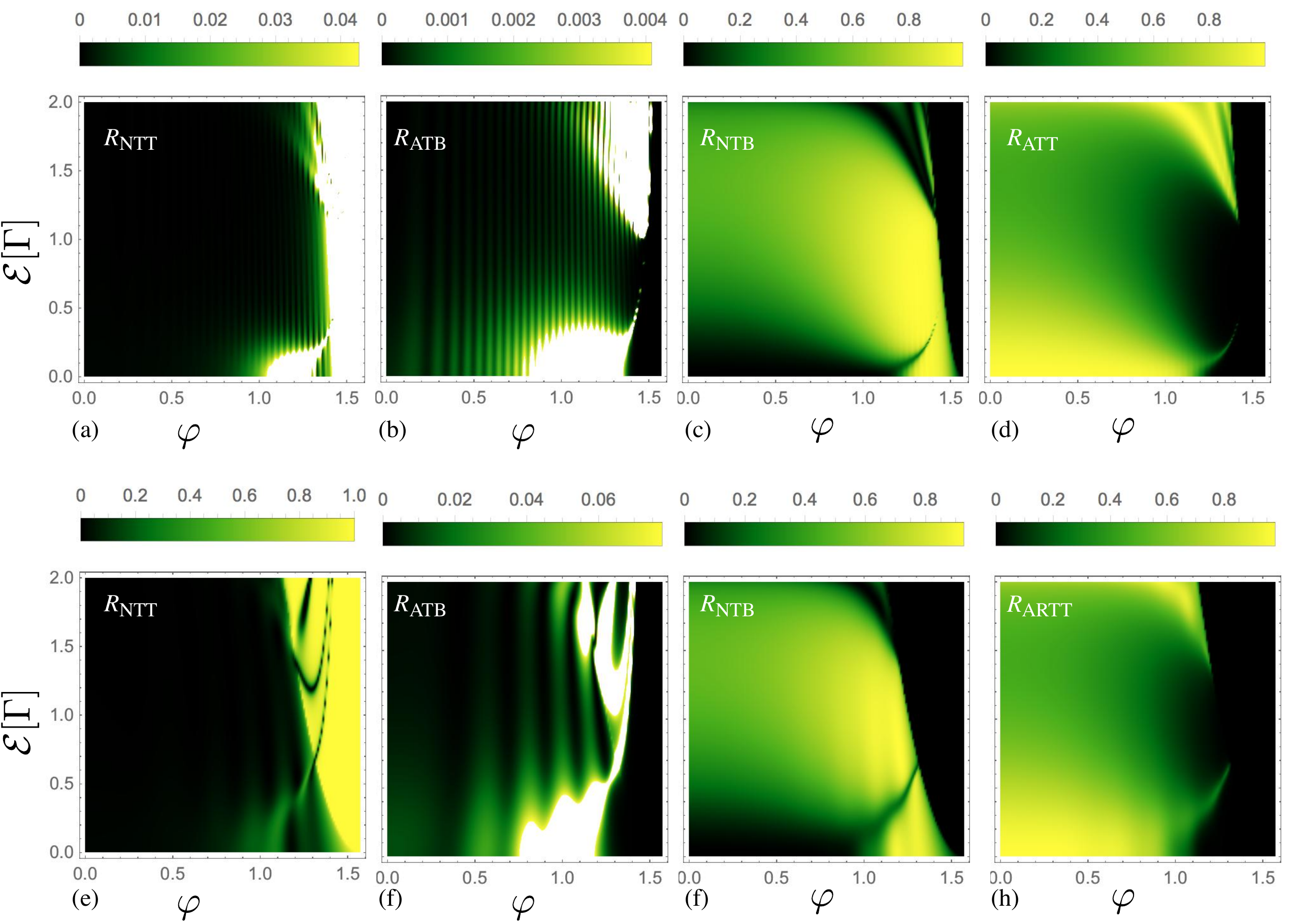}
\caption{\label{reflections} The four reflection channels as a function of injection angle $\varphi$ and injection energy $\mathcal{E}$. Panels (a) to (d), we have set the band overlap to $G=100$, whereas in Panels (e) to (h) we have $G=20$. In all Panels $\Gamma=1$, $\Delta=2\Gamma$ and $L=\xi_\Gamma$.}
\end{center}
\end{figure*}

The momenta in the SSM  [c.f. Fig.~1(d) main text] region are
%
%
%%%%%%%%%%%%
\begin{equation}\label{momenta:SSM}
k_\pm=\sqrt{\frac{2m}{\hbar^2}(G\pm\sqrt{\mathcal{E}^2-\Delta^2})-Q^2}\,.
\end{equation}
%%%%%%%%%%%%
%
%
The coherence factor are the $u_\text{T/B}$ and $v_\text{T/B}$ are the standard one of a $s$-wave superconductor, here T/B is for TL and BL, respectively. Note that there is change of sign between two layers, this is due to their opposite curvature of the energy dispersion:
%
%
%%%%%%%%%%%%
\begin{subequations}\label{coherence:SSM}
\begin{align}
u_\text{T/B} & =\sqrt{\frac{1}{2}\pm\sqrt{1-\left(\frac{\Delta}{\mathcal{E}}\right)}}\\
v_\text{T/B} & =\sqrt{\frac{1}{2}\mp\sqrt{1-\left(\frac{\Delta}{\mathcal{E}}\right)}}
\end{align}
\end{subequations}
%%%%%%%%%%%%
%
%
In the middle EC region, there is a finite Coulomb coupling in the electron-hole bilayer. Here the wave function is given by counter-propagating states for the four possible states: quasi-electron/hole in TL/BL:
%
%
%%%%%%%%%%%%
\begin{widetext}
\begin{align}\label{states:EI}
\psi_\text{EC}(x)& = \ee^{\ii Qy}\left\{\left(a_1 \ee^{\ii q_+ x} +c_1 \ee^{-\ii q_+ x} \right)
\left(\begin{smallmatrix}
u_e \\ v_e \ee^{-\ii \chi} \\0 \\0
\end{smallmatrix}\right)+ \left(b_1 \ee^{-\ii q_- x} +d_1 \ee^{\ii q_- x} \right)
\left(\begin{smallmatrix}
v_e \\ u_e \ee^{-\ii \chi} \\0 \\0
\end{smallmatrix}\right)\right. \nonumber \\
 & \left.+\left(a_2 \ee^{\ii q_+ x} +c_2 \ee^{-\ii q_+ x} \right)
\left(\begin{smallmatrix}
\\0 \\0  \\ v_h \\ u_h \ee^{-\ii \chi}
\end{smallmatrix}\right)+ \left(b_2 \ee^{-\ii q_- x} +d_2 \ee^{\ii q_- x} \right)
\left(\begin{smallmatrix}
\\0 \\0  \\ u_h \\ v_h \ee^{-\ii \chi}
\end{smallmatrix}\right)\right\}\,.
\end{align}
\end{widetext}
%%%%%%%%%%%%
%
%
The momenta in EC region  [c.f. Fig.~1(c) main text] are defined as:
%
%
%%%%%%%%%%%%
\begin{equation}\label{momenta:EI}
q_\pm=\sqrt{\frac{2m}{\hbar^2}(G\pm\sqrt{\mathcal{E}^2-\Gamma^2})-Q^2}\,.
\end{equation}
%%%%%%%%%%%%
%
%
For this case the the coherent function are very similar to the superconducting one, they are here given by:~\cite{Rontani:2005}
%
%
%%%%%%%%%%%%
\begin{subequations}\label{coherence:EI}
\begin{align}
u_{e/h} & = \sqrt{\frac{1}{2}+\sqrt{1-\left(\frac{\Gamma}{\mathcal{E}}\right)}}\\
v_{e/h} & =\sqrt{\frac{1}{2}-\sqrt{1-\left(\frac{\Gamma}{\mathcal{E}}\right)}}
\end{align}
\end{subequations}
%%%%%%%%%%%%
%
%
The density plot of the four reflection channels are presented in S-Fig.~\ref{reflections} for the case of large and small band overlap $G$. It can be observed that for small injection angle $\varphi$ the $R_\text{NTB}$ and the $R_\text{ATT}$ reflection are the dominant processes, on the other side at large angles the $R_\text{NTT}$ and $R_\text{ATB}$ reflection are finite, but smaller than $R_\text{NTB}$ and $R_\text{ATT}$.
%
%
%%%%%%%%%%%%
%%%%%%%%%%%%
%
%

\section{The boundary conditions}

The scattering amplitudes of the various scattering processes can be obtained by imposing the continuity of the wave function in $x=0$ and $x=L$: 
%
%
%%%%%%%%%%%
\begin{subequations}
\begin{align}
\psi_\text{SM}(0^-) & =\psi_\text{EC}(0^+) \\
\psi_\text{EC}(L^-) & =\psi_\text{SSM}(L^+)
\end{align}
\end{subequations}
%%%%%%%%%%
%
%
the continuity of the first derivative in the same points. When  considering a possible mismatch of the Fermi velocity at the interface between the various region, we can model this via delta-barriers of strength $H_\text{SM/EC}$ in $x=0$ and $H_\text{EC/SSM}$ in $x=L$.~\cite{Blonder:1982} In this case the first derivative has a jump that is opposite for the TL and BL~\cite{Rontani:2005}: 
%
%
%%%%%%%%%%%
\begin{subequations}\label{BC}
\begin{align}
\psi_\text{EC}'(0^+)-\psi_\text{SM}'(0^-)=& \frac{2mH_\text{SM/EC}}{\hbar^2} S_z\psi_\text{SM}(0) \label{BC:left}\\ 
\psi_\text{SSM}'(L^+)-\psi_\text{EC}'(L^-)=& \frac{2mH_\text{EC/SSM}}{\hbar^2} S_z \psi_\text{EC}(L)\label{BC:right}\,.
\end{align}
\end{subequations}
%%%%%%%%%%%
%
%
where we have introduced the boundary matrix $S_z$ defined as:
%
%
%%%%%%%%%%%%
\begin{equation}\label{boundary:matrix}
S_z= \mathbb{I}_2 \otimes \sigma_z =\begin{pmatrix}
1 & 0 & 0 & 0 \\
0 & -1 & 0 & 0 \\
0 & 0 & 1 & 0 \\
0 & 0 & 0 & -1
\end{pmatrix}\,.
\end{equation}
%%%%%%%%%%%%
%
%

\section{The critical angle $\alpha_\text{c}$}
%
%
%%%%%%%%%%%%
\begin{figure}[!h]
\begin{center}
\includegraphics[width=0.75\columnwidth]{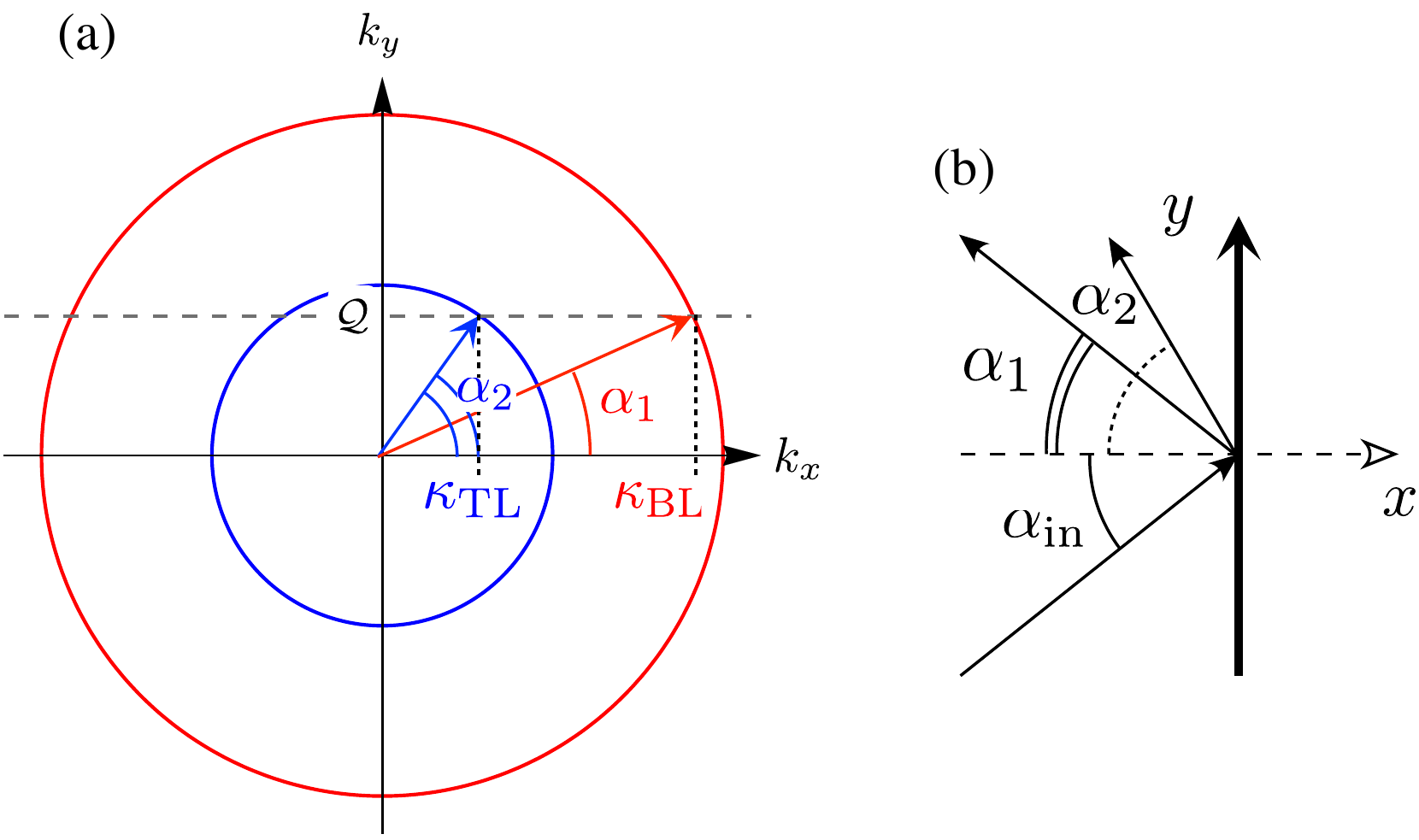}
\caption{\label{fig_c_alpha} Panel (a): The two circle corresponding to the Fermi \emph{surface} for the top  (in red) and the bottom (in blue) layer. The momenta along the $y$-axes $\mathcal{Q}$ is conserved in the scattering process, this produce reflection angles for normal reflection in the TL and for hole reflection in the BL $\alpha_1$, this is opposite to the incoming one $\alpha_\text{in}$. For the case of normal reflection in the BL and hole reflection in the TL, the reflection angle is  $\alpha_2$. Panel (b): a sketch of the reflection angles with respect to the incoming one $\alpha_\text{in}$.}
\end{center}
\end{figure}
%%%%%%%%%%%%
%
%
Using simple kinematic consideration we can determine the expression for the critical angle $\alpha_\text{c}$ Eq.~(4) of the main text. We need only to account for the conservation of the momentum parallel to the interface $k_y=\mathcal{Q}$. For each of the layer involved in the transport mechanism,  momenta in polar coordinate read (see S-Fig.~\ref{fig_c_alpha}):
%
%
%%%%%%%%%%%%
\begin{subequations}\label{momenta:polar}
\begin{align}
k_{\beta,x}&=\kappa_{\beta} \cos\alpha_\text{in}\\
k_{\beta,y}&=\kappa_{\beta} \sin\alpha_\text{in}\\
\end{align}
%%%%%%%%%
%
%
with $\beta\in\{\text{TL,BL}\}$. The moduli are expressed by
%
%
%%%%%%%%%%%%
\begin{align}
\kappa_\text{TL}&= \sqrt{\frac{2m(\mathcal{E}+G)}{\hbar}^2} \\
\kappa_\text{BL}&= \sqrt{\frac{2m(G-\mathcal{E})}{\hbar}^2}\,.
\end{align}
\end{subequations}
%%%%%%%%%%%%
%
%
The normal reflection inside the same layer has a propagation direction opposite to the incoming one, the same is true for electrons that are converted into electrons of the BL.
For electrons injected from the TL and converted into holes of the TL or converted into holes of the BL, the reflection angle is given by:
%
%
%%%%%%%%%%%%
\begin{equation}\label{alpha:special}
\alpha_\frac{\text{ATT}}{\text{NTB}}=\arcsin \left[ \frac{\hbar \mathcal{Q}}{\sqrt{2m(G-\mathcal{E})}}  \right]\,,
\end{equation}
%\end{subequations}
%%%%%%%%%%%%
%
%
for injection energies $\mathcal{E}$ exceeding the band overlap $G$, the angle is becoming complex and the corresponding mode becomes evanescent. By imposing $2m/\hbar^2(G-\mathcal{E})-\mathcal{Q}^2\le 0$ we can determine the critical injection angle $\alpha_\text{c}$:
%
%
%%%%%%%%%%%%
\begin{equation}\label{critical:alpha}
\alpha_\text{c}=\arcsin\left[\sqrt{\frac{|G-\mathcal{E}|}{G+\mathcal{E}}}\right]\,.
\end{equation}
%%%%%%%%%%%%
%
%

\section{Case of a EC region of zero length}
Here we will presents the results in the case the length of the EC region is zero. This corresponds to a lack of a region where carriers of the electron-hole bilayer are coupled. Thus, the hybrid junction reduced to a SM place in contact with a proximized SM. This corresponds to solve the matching problem with wave function \eqref{SM:state} and \eqref{states:SSM}. In the case of the scattering state \eqref{SM:state}, we are considering an initial incoming electron in the TL. Using this initial incoming state, the amplitudes for reflection process read:
%
%
%%%%%%%%%%%%
\begin{subequations}\label{reflectionsSM:SSM}
\begin{align}
r_\text{ATT}&=\frac{u_\text{T} v_\text{T} \ee^{-\ii \phi}}{\gamma} \\
r_\text{NTB}&= 0 \\
r_\text{NTT}&= -\frac{Z (Z-\ii) (u_\text{T}-v_\text{T}) (u_\text{T}+v_\text{T})}{\gamma}\\
r_\text{ATB}&=0\,,
\end{align}
\end{subequations}
%%%%%%%%%%
%
%
whereas for transmissions read:
%
%
%%%%%%%%%%%
\begin{subequations}\label{trasmissionsSM:SSM}
\begin{align}
t_\text{QET}&= \frac{u_\text{T}+\ii u_\text{T} Z}{\gamma}\\
t_\text{QEB}&= 0\\
%\end{align}
%\begin{align}
t_\text{QHT}&=\frac{\ii v_\text{T} Z}{\gamma}\\
t_\text{QHB}&= 0 \,,
\end{align}
\end{subequations}
%%%%%%%%%%%
%
%
here $t_\text{QET/QHT}$ is the transmission as the quasi-electron/quasi-hole in the TL, whereas $t_\text{QEB/QHB}$ is the transmission as the quasi-electron/quasi-hole in the BL.  For both sets of reflection and transmission we have introduced the factor:
%
%
%%%%%%%%%%%%
\begin{equation*}
\gamma=u_\text{T}^2+Z^2 (u_\text{T}-v_\text{T})(u_\text{T}+v_\text{T}) 
\end{equation*}
%%%%%%%%%%%%
%
%
Here $Z=2mH_\text{int}/\hbar^2$ is the dimensionless strength of the delta barrier at the SM/S interface. Importantly, all the interlayer crossing processes have zero probability amplitudes. In the limit of $Z\to 0$ we recover the standard results for S/N junctions.~\cite{Blonder:1982} Starting with a different initial scattering state, \emph{e.g.}, an hole in the BL, we get expressions similar to Eqs.~\eqref{reflectionsSM:SSM}-\eqref{trasmissionsSM:SSM} but still without a direct process between TL and BL.

\section{Analytical results}

In this appendix we show how to get the analytical results (6a)-(6b) of the main text. Both results are obtained assuming that the band overlap is the dominant energy scale $G\gg \max[\mathcal{E},\Gamma,\Delta]$ | as the standard \emph{Andreev-like} approximation in superconducting junctions.~\cite{Blonder:1982} We further assume the superconducting gap is larger that the excitonic one $\Gamma<\Delta$ and injection energy is smaller than superconducting gap : $\mathcal{E}<\Delta$. For zero incidence angle, in these limit we can strongly approximate the values of the momenta in the three regions:
%
%
%%%%%%%%%%%%
\begin{subequations}\label{momenta:Andreev}
\begin{align}
\kappa_\pm& \approx\sqrt{\frac{m}{\hbar^2}}\left(\sqrt{2G}\pm\frac{ \mathcal{E}}{\sqrt{2 G}}\right)=K_0\pm K_\mathcal{E} \label{an:SM:region}\\
q_\pm & \approx \sqrt{\frac{m}{\hbar^2}}\left(\sqrt{2G}\pm \sqrt{\frac{\mathcal{E}^2-\Gamma^2}{2G}}\right)=K_0\pm K_\Gamma\label{an:EI:region}\\
k_\pm & \approx \sqrt{\frac{m}{\hbar^2}}\left(\sqrt{2G}\pm \ii \sqrt{\frac{\Delta^2-\mathcal{E}^2}{2G}}\right)=K_0\pm\ii K_\Delta\label{an:SSM:region}
\end{align}
\end{subequations}
%%%%%%%%%%%%
%
%
For this choice of the energy scale, the coherent function of the superconductor lead $u_\text{T/B}$ and $v_\text{T/B}$ are complex conjugates, thus we need to carry information of only one of them $v_\text{T/B}=u_\text{T/B}^*$ where only $u_\text{T}$ is relevant. Whereas in the EC region we need to keep track of all of them but we can simplify the notation into $u_e=u_h=u$ and $v_e=v_h=v$. We find the coefficients for the components of EC wave function \eqref{states:EI} by solving the boundary problem as first at $x=0$ and at $x=L$, we than impose the equality of the coefficients obtained by at the two different boundaries. By combining all of these we obtain the following sets of equations for the reflection and transmission amplitudes:
%
%
%%%%%%%%%%%%
\begin{widetext}
\begin{eqnarray*}
 t_\text{\text{QET}}2K_{0}uu_\text{T}+t_\text{QHT}(K_{\Gamma}+\ii K_{\Delta})uu_\text{T}^{*}-t_\text{QEB}(K_{\Gamma}+iK_\Delta)vu_\text{T}-t_\text{QHB}2K_{0}vu_\text{T}^{*} = & \\ & \hspace{-5cm}  \ee^{\ii q_{+}L}\left[2K_{0}u+r_\text{NTT}(K_\Gamma-K_\mathcal{E})u-r_\text{NTB}2K_{0}u^{*}\right]\,,
\end{eqnarray*}
\begin{eqnarray*}
 t_\text{QET}(-\ii K_{\Delta}+K_{\Gamma})uu_\text{T}+t_\text{QHT}2K_{0}uu_\text{T}^{*}-t_\text{QEB}2K_{0}vu_\text{T}-t_\text{QHB}(-\ii K_{\Delta}+K_{\Gamma})vu_\text{T}^{*}=& \\ & \hspace{-5cm} \ee^{-\ii q_{+}L}\left[(K_{\Gamma}-K_\mathcal{E})u+r_\text{NTT}2K_{0}u-r_\text{NTB}(K_{\Gamma}+K_\mathcal{E})v\right]\,,
\end{eqnarray*}
\begin{eqnarray*}
t_\text{QET}(\ii K_{\Delta}+K_{\Gamma})vu_\text{T}-t_\text{QHT}2K_{0}vu_\text{T}^{*}+t_\text{QEB}2K_{0}uu_\text{T}-t_\text{QHB}(\ii K_{\Delta}+K_{\Gamma})uu_\text{T}^{*}=& \\ & \hspace{-5cm} \ee^{-\ii q_{_{-}}L}\left[(K_{\Gamma}+K_\mathcal{E})v-r_\text{NTT}2K_{0}v-r_\text{NTB}(K_{\Gamma}-K_\mathcal{E})u\right]\,,
\end{eqnarray*}
\begin{eqnarray*}
 t_\text{QET}2K_{0}vu_\text{T}-t_\text{QHT}(K_{\Gamma}-\ii K_{\Delta})vu_\text{T}^{*}+t_\text{QEB}(K_{\Gamma}-\ii K_{\Delta})uu_\text{T}-t_\text{QHB}2K_{0}uu_\text{T}^{*} =& \\ & \hspace{-5cm} -\ee^{\ii q_{-}L}\left[-2K_{0}v+r_\text{NTT}v(K_\mathcal{E}+K_{\Gamma})+r_\text{NTB}2K_{0}u\right] \,,
\end{eqnarray*}
\begin{eqnarray*}
 t_\text{QET}2K_{0}vu_\text{T}^{*}+t_\text{QHT}(K_{\Gamma}+\ii K_{\Delta})vu_\text{T}-t_\text{QEB}(K_{\Gamma}+\ii K_{\Delta})uu_\text{T}^{*}-t_\text{QHB}2K_{0}uu_\text{T}=& \\ & \hspace{-5cm}  -\ee^{\ii q_{+}L}[(K_{\Gamma}-K_\mathcal{E})u r_\text{ATB}-2K_{0}v r_\text{ATT}]\,,
\end{eqnarray*}
\begin{eqnarray*}
 t_\text{QET}(-\ii K_{\Delta}+K_{\Gamma})vu_\text{T}^{*}+t_\text{QHT}2K_{0}vu_\text{T}-t_\text{QEB}2K_{0}uu_\text{T}^{*}-t_\text{QHB}(-\ii K_{\Delta}+K_{\Gamma})uu_\text{T} = & \\ & \hspace{-2cm} -\ee^{-\ii qL}[2K_{0}u r_\text{ATB}-(K_{\Gamma}+K_\mathcal{E})v r_\text{ATT}]\,, 
\end{eqnarray*}
\begin{eqnarray*}
 t_\text{QET}(\ii K_{\Delta}+K_{\Gamma})uu_\text{T}^{*}-t_\text{QHT}2K_{0}uu_\text{T}+t_\text{QHB}2K_{0}vu_\text{T}^{*}-t_\text{QHB}(\ii K_{\Delta}+K_{\Gamma})vu_\text{T}= & \\ & \hspace{-2cm}  -\ee^{-\ii q_{-}L}[(-K_{\Gamma}+K_{E})u r_\text{ATT}-2K_{0}v r_\text{ATB}]\,,
\end{eqnarray*}
\begin{eqnarray*}
 t_\text{QET}2K_{0}uu_\text{T}^{*}+t_\text{QEB}(-K_{\Gamma}+\ii K_{\Delta})uu_\text{T}-t_\text{QEB}(-K_{\Gamma}+\ii K_{\Delta})u^{*}u_\text{T}^{*}-t_\text{QHB}2K_{0}vu_\text{T} = & \\ & \hspace{-2cm} \ee^{\ii q_{-}L}[(K_{\Gamma}+K_\mathcal{E})v r_\text{ATB}+2K_{0}u r_\text{ATT}]\,.
\end{eqnarray*}

%%%%%%%%%%%%
%
%
We can solve these eight equations with respect to the 4 reflections and transmissions amplitudes we obtain:
%
%
%%%%%%%%%
\begin{subequations}\label{Sebas}
\begin{align}
r_\text{ATT}&=\frac{4\left({\rm Im}[u^{2}]\right)^{2}|u_\text{T}|^{2}}{u_\text{T}^{2}\left(u^{*^{4}}\ee^{-2\delta}+u^{4}\ee^{2\delta}\right)-|u|^{4}\left(u_\text{T}^{*^{2}}4\sinh^{2}\delta+2u_\text{T}^{2}\right)} \nonumber \\ 
 &=\frac{a^{2}\sqrt{1+b^{2}}}{[-a^{2}\cosh(2\delta)-ab\sinh(2\delta)]+i[2b\sinh^{2}(\delta)+a\sinh(2\delta)-ba^{2}]} \label{eq:genRAR2}
\end{align}
 %%%%%%%%
 %
 %
 and
 %
 %
 %%%%%%%%%
\begin{align} 
r_\text{NTB}&=\frac{2\sinh(\delta)|u|^{2}\left[u^{2}\left(u_\text{T}^{*^{2}}\ee^{-\delta}+u_\text{T}^{2}\ee^{\delta}\right)-u^{*^{2}}\left(u_\text{T}^{2}\ee^{-\delta}+u_\text{T}^{*2}\ee^{\delta}\right)\right]}{u_\text{T}^{2}\left(u^{*^{4}}\ee^{-2\delta}+u^{4}\ee^{2\delta}\right)-|u|^{4}\left(u_\text{T}^{*^{2}}4\sinh^{2}\delta+2u_\text{T}^{2}\right)}\nonumber \\ 
& = \frac{4\sinh(\delta)|u|^{2}i\left[a\cosh\delta+b\sinh\delta\right]}{[-a^{2}\cosh(2\delta)-ab\sinh(2\delta)]+i[2b\sinh^{2}(\delta)+a\sinh(2\delta)-ba^{2}]} \label{eq:genRRS2}\,,
\end{align}
\end{subequations}
\end{widetext}
%%%%%%%%%%%
%
%
where in Eqs.~\eqref{eq:genRAR2} and \eqref{eq:genRRS2} we have introduced $a=\sqrt{\Gamma^2-\mathcal{E}^2}/\mathcal{E}$ and $b=\sqrt{\Delta^2-\mathcal{E}^2}/\mathcal{E}$, the constant $\delta=L\xi_\Gamma$ has been introduced in the main text. The other two reflection amplitudes $r_\text{NTT}$ and $r_\text{ATB}$ are smaller than the two in Eqs.~\eqref{Sebas} by an order $1/\sqrt{2G}$ and thus negligible, further, for energies smaller than the superconducting gap $\Delta$ the transmission probabilities are zero. 

If we consider the limit of a long junction $\delta\gg1$ we find that $r_\text{ATT}\to0$ and $r_\text{NTB}\to1$, exactly as we have sound for the full numerical problem in the main text (c.f. Fig.~2 main text).

\section{Case of $\text{HgTe}$ quantum wells}
%
%
%%%%%%%%%%%%
\begin{figure}[!t]
\begin{center}
\includegraphics[width=0.9\columnwidth]{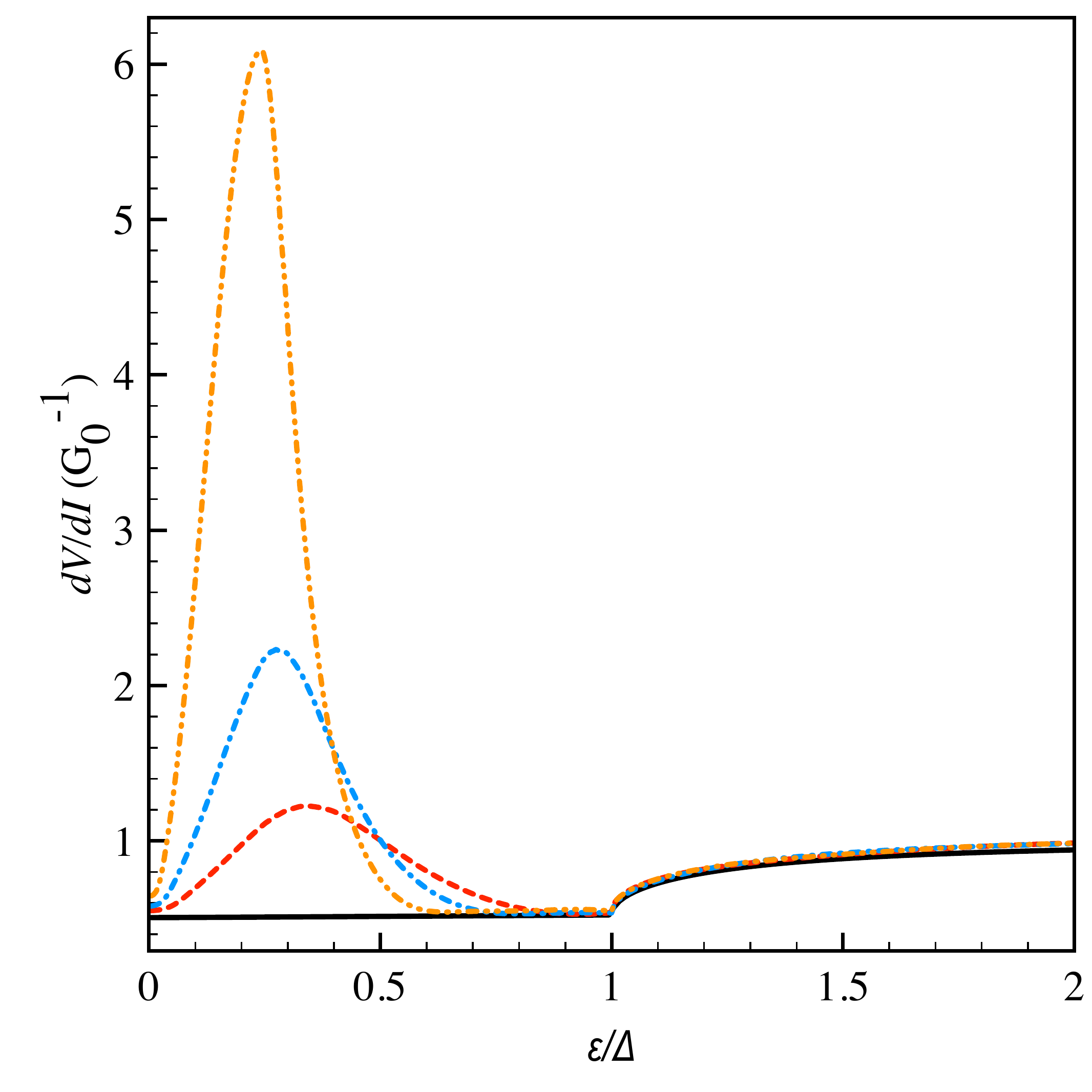}
\caption{\label{figure3sup} Differential resistance as a function of the applied bias and different length of the EC region. In the figure we have set  $\Gamma=1$, $\Delta=12\Gamma$,  $G=60\Gamma$ and $H_\text{SM/EC}=H_\text{EC/SSM}=0.5$, The parameters are chosen in accord to Ref.~[\onlinecite{Kvon:2011,Kononov:2016}]. The different lines refer to $L=0$ (black-solid line), $L=\xi_\Gamma/2$ (red dashed line), $L=0.7\xi_\Gamma$ (blue dashed-dotted line), and $L=\xi_\Gamma$ (orange dashed-dashed-dotted-dotted line).  }
\end{center}
\end{figure}
%%%%%%%%%%%%
%
%
 The experiment by Kononov \emph{et al.},~\cite{Kononov:2016} explored the transport properties of a hybrid junction consisting of  a  HgTe quantum well (QW) with a width of circa 20 nm. From previous studies, the authors know that the system is a indirect SM, with a small bands overlap.~\cite{Kvon:2011} This makes HgTe QWs candidates for observing a direct EC phase.  Specifically in the experiment of Ref.~[\onlinecite{Kononov:2016}], the QW is sandwiched between a normal and a superconductor contact. The transport measurements  showed  a  zero bias anomaly in  the differential resistance with an energy size comparable with the estimated EC gap of the system.~\cite{Kononov:2016} 

We can use the results presented in the main text in order to explain the zero bias anomaly of the hybrid SM/EC/S junction. Assuming that the momentum displacement between the conduction and the valence band is small enough to be negligible, we can evaluate the differential resistance so as explained in the main text. The actual values of the system parameters can be extracted by Refs.~[\onlinecite{Kvon:2011,Kononov:2016}] and have: $G=3.0$~meV, $\Gamma=0.05$~meV and $\Delta=0.6$~meV, which should correspond to a (Nb/FeNb)/HgTe structure explored  in  Ref.~[\onlinecite{Kononov:2016}], we have further set a finite transparency of the two interface barriers $H_\text{L}=H_\text{R}=0.5$.~\cite{Kononov:2016} For these parameters value we have an EC coherence length of $\xi_\Gamma=955$~nm. 

The results for the differential resistance are shown in S-Fig.~\ref{figure3sup}. The black solid line describes a SM/SSM junction ($L=0$), and coincides with the well-known results of the BTK theory after integration over injection angle (main text).~\cite{Mortensen:1999} It is important to  note that $\Delta/G=0.50$ and hence the differential resistance at zero voltage is not exactly equal to $0.5G_0^{-1}$, which would be the result  obtain in the leading order when $\Delta/G\ll1$.  
By increasing the length of the EC region we obtain  peak in the differential resistance at energies smaller than $\Delta$.  This result resemble the  experimental observations of  Ref.~[\onlinecite{Kononov:2016}] that has been attributed as a manifestation  of the  EC phase. 

The experiment of   Ref.~[\onlinecite{Kononov:2016}] also shown features  in  the  superconducting sub-gap conductance that resemble the multiple-Andreev reflections (MARS)  processes in  voltage biased Josephson junctions.~\cite{Averin:1995}. In principle  such  sub-gap  features are  unexpected, since in  the experiment by Kononov \emph{et al.}~\cite{Kononov:2016} there is only  one superconducting electrode. 
Our model is also done for an unique S-electrode and in a quasi-equilibrium situation and  therefore cannot describe MARS-like processes.  

\end{document}